# Solution of the Skyrme-Hartree-Fock equations in the Cartesian deformed harmonic-oscillator basis. (III) HFODD (v1.75r): a new version of the program.

J.Dobaczewski[a,b,1] and J. Dudek[a,2]

[a]Institute de Recherches Subatomiques, CNRS-IN$_2$P$_3$/Université Louis Pasteur,
F-67037 Strasbourg Cedex 2, France
[b]Institute of Theoretical Physics, Warsaw University
ul. Hoża 69, PL-00681 Warsaw, Poland

**Abstract**

We describe the new version (v1.75r) of the code HFODD which solves the nuclear Skyrme-Hartree-Fock problem by using the Cartesian deformed harmonic-oscillator basis. Three minor errors that went undetected in the previous version have been corrected. The new version contains an interface to the LAPACK subroutine ZHPEV. Several methods of terminating the Hartree-Fock iteration procedure, and an algorithm that allows to follow the diabatic configurations, have been implemented.

## NEW VERSION PROGRAM SUMMARY

*Title of the program:* HFODD (v1.75r)

*Catalogue number:*

*Program obtainable from:* CPC Program Library, Queen's University of Belfast, N. Ireland (see application form in this issue)

*Reference in CPC for earlier version of program:* J. Dobaczewski and J. Dudek, Comput. Phys. Commun. **102** (1997) 183 (v1.60r).

*Catalogue number of previous version:* ADFL

*Licensing provisions:* none

*Does the new version supersede the previous one:* yes

*Computers on which the program has been tested:* CRAY C-90, SG Power Challenge L, IBM RS/6000, Pentium-II, Athlon

*Operating systems:* UNIX, UNICOS, IRIX, AIX, LINUX

---

[1]E-mail: jacek.dobaczewski@fuw.edu.pl
[2]E-mail: jerzy.dudek@ires.in2p3.fr

*Nature of physical problem*
The nuclear mean-field and an analysis of its symmetries in realistic cases are the main ingredients of a description of nuclear states. Within the Local Density Approximation, or for a zero-range velocity-dependent Skyrme interaction, the nuclear mean-field is local and velocity dependent. The locality allows for an effective and fast solution of the self-consistent Hartree-Fock equations, even for heavy nuclei, and for various nucleonic ($n$-particle $n$-hole) configurations, deformations, excitation energies, or angular momenta.

*Method of solution*
The program uses the Cartesian harmonic oscillator basis to expand single-particle wave functions of neutrons and protons interacting by means of the Skyrme effective interaction. The expansion coefficients are determined by the iterative diagonalization of the mean field Hamiltonians or Routhians which depend nonlinearly on the local neutron and proton densities. Suitable constraints are used to obtain states corresponding to a given configuration, deformation or angular momentum. The method of solution has been presented in: J. Dobaczewski and J. Dudek, Comput. Phys. Commun. **102** (1997) 166.

*Summary of revisions*

1. An error in the calculation of one of the time-odd mean-field potentials has been corrected.
2. A factor in the calculation of the multipole moment $Q_{22}$ has been corrected.
3. Scaling of the coupling constants has been corrected.
4. An interface to the LAPACK subroutine ZHPEV has been created.
5. Several methods of terminating the Hartree-Fock iteration procedure have been implemented.
6. An algorithm that allows to follow the diabatic configurations has been implemented.
7. Saving of auxiliary data for a faster calculation of the Coulomb potential has been implemented.

8. Calculation of average quadrupole moments and radii of single-particle states has been added.
9. Calculation of the Bohr deformation parameters has been added.

*Restrictions on the complexity of the problem*
The main restriction is the CPU time required for calculations of heavy deformed nuclei and for a given precision required. One symmetry plane is assumed. Pairing correlations are only included in the BCS limit and for the conserved time-reversal symmetry (i.e., for non-rotating states in even-even nuclei).

*Typical running time*
One Hartree-Fock iteration for the superdeformed, rotating, parity conserving state of $^{152}_{66}$Dy$_{86}$ takes about nine seconds on the CRAY C-90 computer. Starting from the Woods-Saxon wave functions, about fifty iterations are required to obtain the energy converged within the precision of about 0.1 keV. In case when every value of the angular velocity is converged separately, the complete superdeformed band with precisely determined dynamical moments $\mathcal{J}^{(2)}$ can be obtained within one hour of CPU on the CRAY C-90, or within three hours of CPU on the Athlon-550 MHz processor. This time can be often reduced by a factor of three when a self-consistent solution for a given rotational frequency is used as a starting point for a neighboring rotational frequency.

*Unusual features of the program*
The user must have an access to the NAGLIB subroutine F02AXE or to the ESSL or LAPACK subroutine ZHPEV which diagonalize complex hermitian matrices, or provide another subroutine which can perform such a task. The LAPACK subroutine ZHPEV can be obtained from the Netlib Repository at University of Tennessee, Knoxville: http://netlib2.cs.utk.edu/cgi-bin/netlibfiles.pl?filename=/lapack/complex16/zhpev.f

LONG WRITE-UP

# 1 Introduction

The method of solving the Hartree-Fock (HF) equations in the Cartesian harmonic oscillator (HO) basis was described in the previous publication, Ref. [1], which is below referred to as I. The previous version of the code HFODD (v1.60r) was published in Ref. [2] which is below referred to as II. The present paper is a long write-up of the new version (v1.75r) of the code HFODD. This extended version is fully compatible with all previous versions.

Information provided in I and II remains valid, unless explicitly mentioned in the present long write-up.

In Section 2 we briefly review the modifications introduced in version (v1.75r) of the code HFODD. In particular, in Section 2.6 we present numerical tests of the code in situations when the iteration procedure does not converge because of a crossing of single-particle levels in function of the rotational frequency. We then introduce a procedure to follow configurations along the so-called diabatic path. Such a procedure was implemented in version (v1.75r), and allows, in many cases, to obtain converged solutions for both crossing configurations.

Section 3 lists all additional new input keywords and data values, introduced in version (v1.75r). The structure of the input data file remains the same as in the previous versions, see Section 3 of II. Similarly, all previously introduced keywords and data values retain their validity and meaning.

## 2  Modifications introduced in version (v1.75r)

### 2.1  Calculation of the density $\Delta s$

There are two misprints in formulas (I-46)[3] for the Cartesian components of the vector density $\Delta s$. The correct expressions are as follows:

$$\Delta s_1 = 2\Re\left(L^{+-} + L^{-+}\right) + 2T_1, \tag{1a}$$

$$\Delta s_2 = -2\Im\left(L^{+-} - L^{-+}\right) + 2T_2, \tag{1b}$$

$$\Delta s_3 = 2\Re\left(L^{++} - L^{--}\right) + 2T_3. \tag{1c}$$

The misprints were present only in the text of I, and did not affect the code HFODD (v1.60r). Unfortunately, the factors of 2, which should multiply densities $T_1$ and $T_2$ in Eqs. (1a) and (1b), respectively, where missing in version (v1.60r) of the code. As far as the numerical values are concerned, this error has been fairly unimportant for the final results, however, it may have created a weak spurious dependence of the results on the orientation of the nucleus with respect to the Cartesian reference frame, because only the $x$ and $y$ components of $\Delta s$ were affected.

Incorrect expressions for the densities $\Delta s_1$ and $\Delta s_2$ amounted to adding the erroneous term $-C_t^{\Delta s}\left(s_{xt}T_{xt} + s_{yt}T_{yt}\right)$ to the time-odd energy density $\mathcal{H}_t^{\rm odd}(\boldsymbol{r})$ of Eq. (I-12a), and simultaneously adding the erroneous terms $-2C_t^{\Delta s}T_{xt}$ and $-2C_t^{\Delta s}T_{yt}$ to the time-odd spin potentials $\Sigma_{xt}$ and $\Sigma_{yt}$, respectively, Eq. (I-29b). Therefore, up to a very small difference between the matrix elements of spin and kinetic-spin potentials, the incorrect densities $\Delta s_1$ and $\Delta s_2$ where equivalent to a (direction-dependent) modification of the coupling constant $C_t^T$. This is the main reason why the error went undetected for a relatively long time. Needless to say, calculations performed with $C_t^{\Delta s}=0$, and in particular those with all time-odd terms neglected, are unaffected.

Since the term $\boldsymbol{s}_t \cdot \Delta\boldsymbol{s}_t$ gives anyhow fairly small contribution to the rotational properties of nuclei (compare curves denoted by open circles with those denoted by full squares in Figs. 3 and 4 of Ref. [3]), the incorrect expressions for densities $\Delta s_1$ and $\Delta s_2$ had numerically relatively small importance. The total energies could have been affected at the level of about $0.3\,\text{MeV}$ and the total spins at the level of about $0.3\,\hbar$ (for details compare the output file reproduced in the section TEST RUN OUTPUT below with that given in II).

---

[3]Symbol (I-46) refers to Eq. (46) of I.

## 2.2 Calculation of the multipole moment $Q_{22}$

In version (v1.60r) of the code HFODD, a factor of $\sqrt{6}$ was missing in the values of the multipole moment $Q_{22}$ printed on the output file.

Values of the quadrupole moments $Q_{20}$ and $Q_{22}$ depend on the normalization factors which in the standard nuclear-physics applications are different than those used in electrodynamics, cf. Ref. [4]. In particular, the code HFODD (v1.75r) uses the normalization factors such that

$$Q_{20} = \langle 2z^2 - x^2 - y^2 \rangle, \qquad (2a)$$
$$Q_{22} = Q_{20} \times \tan\gamma, \qquad (2b)$$

where $\gamma$ is the standard angle measuring the non-axial quadrupole deformation [5].

## 2.3 Scaling of the coupling constants

As described in Section 3.2 of II, the code HFODD allows for the scaling of the Skyrme-functional coupling constants by arbitrary factors which are read from the input data file. The scaling can be independently performed in the total-sum and isoscalar-isovector representations of the coupling constants, see Eqs. (I-14) and (I-15).

In version (v1.60r), the scaling formulas were incorrectly coded and as a result, under some very special circumstances described below, calculations were performed for values of the scaled coupling constants which were different from those intended by the user. This was happening provided that for the given coupling constant two conditions occurred simultaneously:

1. Scaling factor of the isovector coupling constant was different than 1.

2. Scaling factor of the isovector coupling constant was different than that of the isoscalar coupling constant.

This is a fairly unusual combination of scaling conditions, and hence the error escaped detection for a relatively long time. In particular, all results obtained in Ref. [3] are correct because they were obtained with equal scaling factors of the isovector and isoscalar coupling constant, and hence condition 2. above was not fulfilled. Obviously, calculations performed without scaling, i.e., with all scaling factors equal 1, are unaffected.

## 2.4 Diagonalization subroutines

As described in Section 5.1, an interface to the LAPACK subroutine ZHPEV has been created. This allows using public-domain diagonalization subroutines, as an alternative of using the NAGLIB or ESSL packages, previously described in Section 5.3 of II.

## 2.5 Terminating the Hartree-Fock iteration procedure

As described in Section 3.1, the code HFODD (v1.75r) can detect several conditions which allow to stop the iteration procedure before the predefined number of iterations is exhausted. This includes stopping after the convergence is reached, and a required stability of the solutions is obtained, as well as an early stopping of the iteration procedure which is not going to converge.

## 2.6 Diabatic blocking

Crossing or non-crossing ("repulsion") of the mean-field single-particle energy levels in function of some continuous parameter(s) of theory, such as the rotational frequency and/or the constraining multipole moments, is one of the most important phenomena in the studies of high-spin nuclear states. They may give rise to several kinds of characteristic irregularities occurring along the rotational bands and these irregularities often help significantly in assigning the theoretical single-particle configurations to the experimental rotational bands.

As it is well known, levels that differ in terms of their discrete good quantum numbers such as, for instance, simplex, parity, signature ... etc., will generally cross. These crossings are such that the underlying intrinsic characteristics (e.g. single-particle alignments or multipole moments) do not change in any remarkable way before and after the crossing and this mechanism will be of no interest here.

In contrast, levels that belong to the same discrete symmetry will most of the time approach each other in function of the parameter studied and then go apart but in such a way that the intrinsic characteristics of the first one will go over to the second one and *vice versa*. This non-crossing rule, sometimes called the Landau-Zener effect, cannot be given any more rigorous general formulation, and it may happen that the distance of the closest approach for the same symmetry levels is zero. In those, in practice very rare cases, the two crossing mechanisms mentioned do not differ.

An example of a crossing of two HF configurations (no pairing) is presented in Fig. 1. The left panel shows the total energies $E(I)$ as functions of the total spin $I$, while the right panel shows the total Routhians $R(\omega)=E(\omega)-I(\omega)\omega$ as functions of the rotational frequency $\hbar\omega$. Since both quantities vary very rapidly as functions of their arguments, they are plotted with respect to the corresponding rigid-rotor reference values, i.e., $E(I)$ is shown relatively to $I(I+1)/(2J_0)$, and $R(\omega)$ is shown relatively to $-J_0\omega^2/2$, where the constant rigid-rotor moment of inertia of $J_0=100\,\hbar^2$/MeV has been used. Calculations have been performed within the cranking approximation for values of the rotational frequency of $\hbar\omega$=0.5(0.05)0.8 MeV.

The examples shown in Fig. 1 correspond to two bands in $^{151}$Tb (see Ref. [6] for a more complete description of calculations performed in this nucleus). Both configurations contain the same set of the single-particle levels of $^{150}$Tb being occupied. They correspond to the neutron and proton configurations of $(N_{++}, N_{+-}, N_{-+}, N_{--})$=(22,21,21,21) and (15,16,17,17), respectively, as described by the numbers of states occupied in the parity-signature blocks $(\pi,r)$=(+1,+i), (+1,−i), (−1,+i), and (−1,−i). The ground-state band in $^{151}$Tb can be obtained by putting the 86th neutron into the lowest available level for $N_{+-}$=22, thus obtaining the closed $N$=86 SD magic neutron configuration. Since the order of orbitals may change with changing rotational frequency, numbers $N_{\pi,-ir}$ are not necessarily the most practical for defining physical characteristics of the single-particle states in question. Usually, one uses the so-called asymptotic Nilsson quantum numbers $[Nn_z\Lambda]\Omega$ [5] for that purpose. Code HFODD calculates these quantum numbers by finding the dominant Nilsson components of the HF single-particle states. Several excited bands in $^{151}$Tb can be obtained by putting the 86th neutron into one of the higher available levels, for instance, in the $N_{-+}$=22, 23, or 24 levels, that correspond to one of the [521]3/2($r=+i$), [514]9/2($r=+i$), or [761]3/2($r=+i$) Nilsson orbitals.

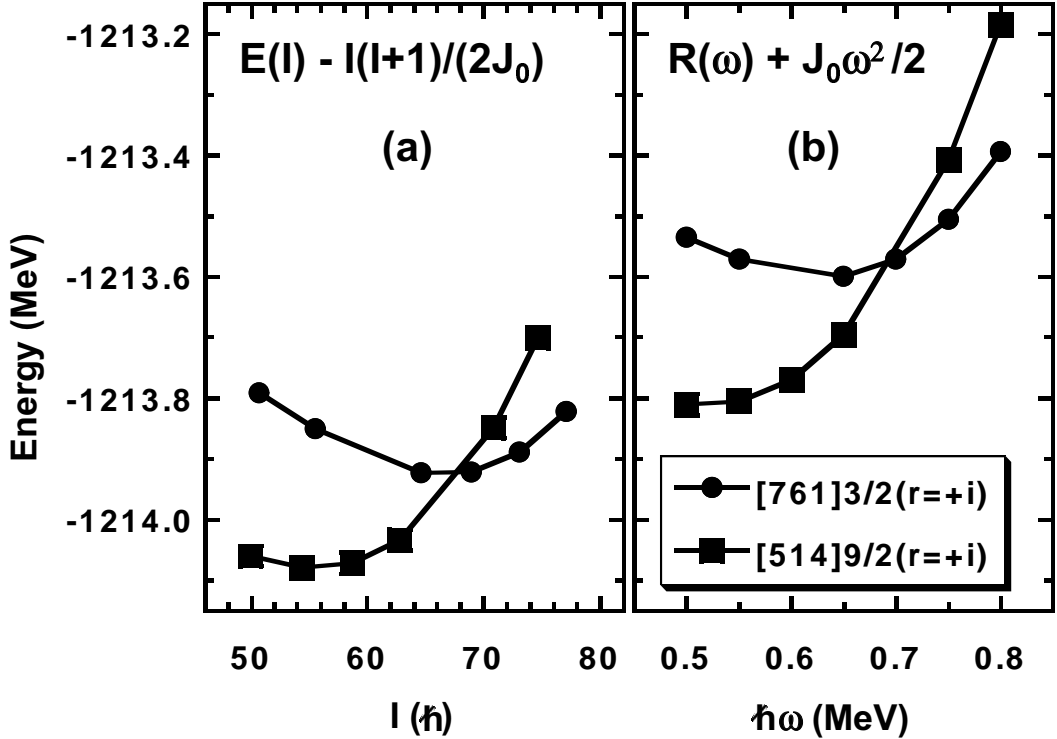

Figure 1: Total energies $E(I)$, (a), and total Routhians $R(\omega)$, (b), of the [761]3/2($r=+i$) and [514]9/2($r=+i$) neutron diabatic configurations in $^{151}$Tb. The rigid-rotor reference energies for $J_0$=100 $\hbar^2$/MeV have been subtracted. (In our graphical representation we employ the convention according to which the levels that carry the smoothly varying intrinsic characteristics (see text) are denoted with the same symbols. This is at variance with the convention that stresses the Landau-Zener mechanism, as used by some authors. According to the latter one, the sequences of the lowest-lying points would have been drawn as squares and that of the higher-lying points as circles.)

Of particular interest in $^{151}$Tb are exited bands in which the [761]3/2($r=+i$) intruder orbital is occupied. This orbital decreases in energy with increasing rotational frequency, and therefore, it crosses the other ($\pi,r$)= ($-1,+i$) orbitals. In particular, in $^{151}$Tb the [761]3/2($r=+i$) orbital corresponds to the $N_{-+}$=24th orbital at low frequencies, then it becomes the $N_{-+}$=23rd orbital, and finally, at high frequencies it is the lowest available $N_{-+}$=22nd orbital. Configurations shown in Fig. 1 correspond to the crossing of the [761]3/2($r=+i$) and [514]9/2($r=+i$) orbitals. Following the standard convention, these configurations are called the diabatic ones, because they correspond to the given orbital being occupied, irrespective of its excitation energy. On the other hand, configurations based on occupying the $N_{-+}$=22, 23, or 24 states are called the adiabatic ones. Obviously, in adiabatic configurations, different Nilsson orbitals are occupied at different frequencies.

Fig. 2 shows the negative-parity single-particle neutron Routhians in $^{151}$Tb. The left and right panels show the results obtained for the $[761]3/2(r=+i)$ and $[514]9/2(r=+i)$ diabatic configurations, respectively. The occupied orbitals are in both cases denoted by the filled symbols. It is clear that the crossing frequency depends on which of the shown orbitals is occupied. This is due to the self-consistent effects that influence the deformations, spins, and other characteristics of many-body states, calculated at given rotational frequencies and for given particle-hole configurations.

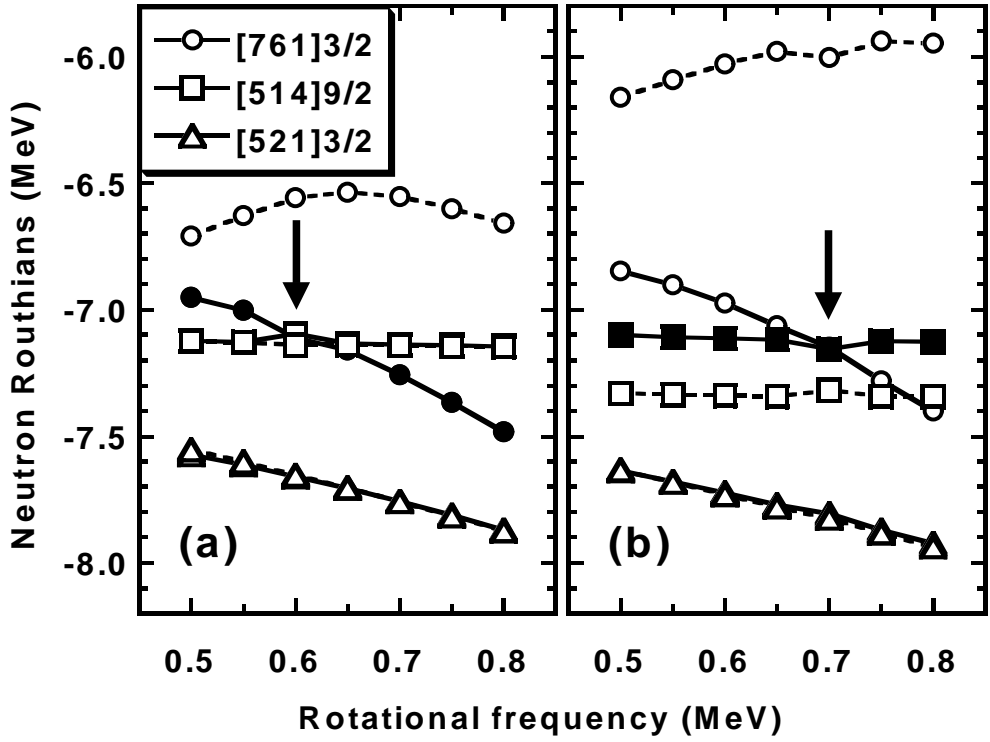

Figure 2: Negative-parity neutron single-particle Routhians in $^{151}$Tb calculated for the $[761]3/2(r=+i)$ (a) and $[514]9/2(r=+i)$ (b) neutron diabatic configurations. Solid and dashed curves denote the $r=+i$ and $r=-i$ signatures, respectively. The arrows denote the angular frequencies where the converged solutions near the crossing points could not be found.

In the calculations, one has to begin the analysis by finding the adiabatic configurations, in order to find the lowest available orbitals, to determine their physical characteristics, and to check whether or not there are any crossings of configurations that should be followed diabatically. The input data files `tb151-a.dat` and `tb151-b.dat` give examples of runs that find the adiabatic configurations for the $N_{-+}$=23rd and $N_{-+}$=24th orbitals being occupied, respectively.

After several HF iterations, the levels of the same symmetry, that are energetically sufficiently far, change their intrinsic characteristics (such as, e.g., angular momentum alignments) only by very small amounts. This is not true for the case of the crossing when the two states in question exchange their relative positions from one iteration to another. Therefore, in the HF calculations, very often the Landau-Zener avoided crossings of levels manifest themselves in the form of diverging iteration procedure. This observation is used by the code HFODD (v1.75r) in order to locate the crossing by recognizing the oscillatory ("ping-pong") behavior of solutions.

Table 1: Example of the output printed when the "ping-pong" divergence is found, see text.

```
********************************************************************************
*                                                                              *
*    TABLE BELOW GIVES THE MAXIMUM VALUES OF DIFFERENCES OF ALIGNMENTS BETWEEN  *
*    THE LAST TWO ITERATIONS. IT MAY SERVE AS A GUIDE TO PIN DOWN STATES WHICH  *
*    EXCHANGE WAVE FUNCTIONS IN EVERY SECOND ITERATION. THEY MAY BE CANDIDATES  *
*    F O R    T H E    D I A B A T I C    B L O C K I N G    C A L C U L A T I O N  *
*                                                                              *
********************************************************************************
*                    *                                                         *
*       BLOCKS       * ANGULAR MOMENTUM ALIGNMENT      INTRINSIC SPIN ALIGNMENT *
*       ------       * ---------------------------    --------------------------*
*                    *    PARTICLES       HOLES           PARTICLES       HOLES*
*  ISO. PAR. SIG.    *  INDEX VALUE   INDEX VALUE       INDEX VALUE   INDEX VALUE *
*                    *                                                         *
********************************************************************************
*                    *                                                         *
* NEUT   +    +      *    38   0.221    13   0.018       24   0.023    13   0.016 *
* NEUT   +    -      *    28   0.163     7   0.041       33   0.014     7   0.032 *
* NEUT   -    -      *    39   0.095    13   0.198       44   0.015    14   0.150 *
* NEUT   -    +      *    24   2.820    23   2.807       24   0.164    23   0.167 *
* PROT   +    +      *    24   0.255    13   0.149       25   0.129    14   0.122 *
* PROT   +    -      *    27   1.733    13   0.062       21   0.319     8   0.020 *
* PROT   -    -      *    32   0.119    10   0.530       32   0.029    10   0.065 *
* PROT   -    +      *    24   2.649    15   0.082       24   0.153    10   0.030 *
*                    *                                                         *
********************************************************************************
*                                                                              *
*    FOLLOWS THE LIST OF CANDIDATE CONFIGURATIONS FOR THE DIABATIC BLOCKING    *
*                                                                              *
********************************************************************************
*                    *                                                         *
* NEUT   -    +      *    24   2.820    23   2.807                             *
* NEUT   -    +      *                                    24   0.164    23   0.167 *
*                    *                                                         *
********************************************************************************
```

The "ping-pong" divergence is a result of the standard self-consistent prescription, for which the occupied states in a given iteration are the eigenstates of the mean-field Hamiltonian diagonalized in the preceding iteration. Such a divergence is characterized by two series of HF states, each one appearing in every second HF iteration. Focusing our attention on the specific example discussed above, one can describe such a divergence in the following way. Suppose that in a given HF iteration, the single-particle orbital $[761]3/2(r=+i)$ appears below the $[514]9/2(r=+i)$ orbital, i.e., it corresponds to $N_{-+}$=23. In this particular iteration it becomes occupied, and is included in all densities, for which the mean-field Hamiltonian is determined. In the next iteration, the new Hamiltonian is diagonalized, however, at present the orbital [514]9/2($r$=+$i$) appears below [761]3/2($r$=+$i$), and it is this orbital, again for $N_{-+}$=23, which becomes occupied. It is obvious that such a situation may repeat itself infinitely many times, never leading to a self-consistent solution.

In many cases, it is possible to find converged solutions near the crossing points simply by decreasing the convergence rate, i.e., by using parameters `SLOWEV` and `SLOWOD` (see Sec. 3.5 of II) only slightly smaller then 1. This method works well provided that the two crossing orbitals have fairly different characteristics, like it is the case in the example discussed here. However, in the case of a strong mixing between the two crossing orbitals, the HF iteration procedure converges near the crossing points only seldomly.

Upon recognizing the "ping-pong" divergence condition (see Sec. 3.1 for details of the procedure), the code HFODD (v1.75r) prints the summary table that may help in identifying the crossing orbitals. For the $N_{-+}$=23 adiabatic configuration at $\hbar\omega$=0.65 MeV, the table has the form presented in Table 1.

In the upper part of the table, the code prints the absolute values of differences $|\Delta i_y|$ ($|\Delta s_y|$) of single-particle average alignments $i_y$ (intrinsic spins $s_y$) between the last and the last but one iterations (columns denoted by `VALUE`). The results are printed separately for the particle (empty) and hole (occupied) states, and their indices are given in columns denoted by `INDEX`. In each of the charge-parity-signature (or charge-simplex) blocks the results are printed only for those states for which the values of $|\Delta i_y|$ and $|\Delta s_y|$ are the largest ones.

In the lower part of the table, the code prints the list of particle-hole pairs from the upper part of the table, for which the indices (`INDEX`) differ by one, i.e., which correspond to the particle-hole pairs at the Fermi surface. In the specific example discussed here, the code properly identifies the crossing of the $N_{-+}$=23 and $N_{-+}$=24 orbitals, and proposes the diabatic blocking based on the average values of the single-particle alignments, see Sec. 3.2. The crossing orbitals, [761]3/2($r$=+$i$) and [514]9/2($r$=+$i$), have markedly different alignments of about $i_y$=+2.23 and $-0.46\,\hbar$, respectively. This corresponds to values of $|\Delta i_y|$=+2.820 or $2.807\,\hbar$ printed in the table. Therefore, by requesting that the state that has a larger alignment be occupied (among the $N_{-+}$=23 and $N_{-+}$=24 orbitals), one obtains the diabatic [761]3/2($r$=+$i$) configuration. Similarly, by choosing the state that has a smaller alignment, one obtains the [514]9/2($r$=+$i$) diabatic configuration, and this is so irrespective of which orbital has smaller or larger energy, i.e., irrespective of the angular frequency. The input data files `tb151-c.dat` and `tb151-d.dat` give examples of runs that find the diabatic configurations for the [761]3/2($r$=+$i$) and [514]9/2($r$=+$i$) orbitals being occupied, respectively.

## 2.7 Saving the Coulomb data

As described in Section 3.3, the code HFODD (v1.75r) can save on disk auxiliary data for a faster calculation of the Coulomb potential. These data can be reused in subsequent runs of the code, provided the Coulomb parameters `NUMCOU`, `NUMETA`, and `FURMAX` (see Section 3.5 of II) do not change from one run to the next.

## 2.8 Calculation of moments and radii

Average quadrupole moments, $\langle q_{20}\rangle$ and $\langle q_{22}\rangle$, radii squared, $\langle r^2\rangle$, and sizes squared, $\langle x^2\rangle$, $\langle y^2\rangle$, and $\langle z^2\rangle$, of single-particle states are calculated and stored in the review file `FILREV`, see Section 3.9 of II.

## 2.9 Calculation of the Bohr deformation parameters

In version (v1.75r), the Bohr deformation parameters $\alpha_{\lambda\mu}$ [5] are calculated using the first-order approximation. The code uses the simplest linearized expression (cf. Eq. (1.35) of Ref. [5]) relating deformation parameters to multipole moments of a sharp-edge uniformly charged shape, i.e.,

$$\alpha_{\lambda\mu} = \frac{4\pi\sqrt{1+\delta_{\mu 0}}}{3NR_0^\lambda}\langle r^\lambda Y_{\lambda\mu}\rangle. \tag{3}$$

For $\mu\neq 0$ the deformation parameters contain the standard factor of $\sqrt{2}$. This reflects the fact that for the assumed $\hat{S}_y$ simplex symmetry (see Sec. 3.1 of I), the values of multipole moments for $\mu<0$ are up to a phase equal to those for $\mu>0$. Neutron, proton, or mass deformation parameters are calculated from the corresponding neutron, proton, or mass multipole moments $\langle r^\lambda Y_{\lambda\mu}\rangle$, and by setting $N$ in Eq. (3) equal to the number of neutrons, protons, or nucleons. The equivalent radii $R_0$ are respectively calculated from the neutron, proton, or mass rms radii as $\sqrt{5/3}\langle r^2\rangle^{1/2}$. One should note that for large deformations the neglected higher-order terms (see Eq. (1.35) of Ref. [5]) will in general be non-negligible. Therefore, the printed values of the Bohr deformation parameters should be used only as (often very) rough estimates.

## 3 Input data file

Structure of the input data file has been described in II, and in version (v1.75r) of the code HFODD this structure is exactly the same. All previous items of the input data file remain valid, and several new items were added, as described in Secs. 3.1–3.3.

Together with the FORTRAN source code in the file `hfodd.f`, several examples of the input data files are provided. File `dy152-c.dat` contains all the valid input items, and the input data are identical to the default values. Therefore, the results of running the code with the input data file `dy152-c.dat` are identical to those obtained for the input data file containing only one line with the keyword `EXECUTE`.

File `dy152-d.dat` contains all those input items that were added in version (v1.75r), and the input data there are those recommended for being used in the new version. This file is reproduced in the section TEST RUN INPUT below.

### 3.1 General data

*Keyword:* `ITERAT_EPS`
$$0.0 = \text{EPSITE}$$
*Keyword:* `MAXANTIOSC`
$$1 = \text{NULAST}$$

These two parameters govern the termination of the HF iterations according to the achieved stability of solutions. The stability of the HF energy has been defined in (I-37) as the difference between the total energies calculated from the single-particle energies and from the Skyrme functional. The HF iterations continue until the absolute value of the stability is smaller than `EPSITE` (in MeV) over `NULAST` consecutive iterations. When this condition is fulfilled, iteration procedure terminates and the final results are printed. This allows for an automated adjustment of the number of iterations which are required to achieve a given level of convergence. The number of iterations `NOITER`, Section 3.1 of II, can now be set to a large value at which the iterations terminate if a stable solution is not found.

The default value of `EPSITE`=0.0 ensures that whenever this new option is not used, the code HFODD (v1.75r) behaves as that in version (v1.60r). If a non-zero value of `EPSITE` is used, a non-zero value of `NULAST` has to be used too. In practice, a value of `NULAST`=5 prevents an accidental termination of iterations in all cases when the stability energy changes the sign, but the solution is not yet self-consistent.

*Keyword:* `PING-PONG`

$$0.0, 3 = \text{EPSPNG, NUPING}$$

The code is able to detect the "ping-pong" divergence described in Section 2.6, i.e., the situation when the HF iteration procedure gives alternating results in every second iteration. Upon continuing the iteration, both sequences of results, i.e., those which correspond to the iteration numbers being even and odd, stay different but perfectly stable, and hence the correct self-consistent solution is never attained.

The code recognizes such a situation by calculating the averages and variances of the stability energy (I-37), separately in the even and in the odd sequences of results, over the last `NUPING` pairs of iterations. The "ping-pong" divergence condition occurs when both variances become a factor `EPSPNG` smaller than the absolute value of the difference of the corresponding averages, i.e., when

$$\Delta(\delta\mathcal{E})_{\text{even}} < \text{EPSPNG} \times |\overline{\delta\mathcal{E}}_{\text{even}} - \overline{\delta\mathcal{E}}_{\text{odd}}|, \tag{4a}$$

$$\Delta(\delta\mathcal{E})_{\text{odd}} < \text{EPSPNG} \times |\overline{\delta\mathcal{E}}_{\text{even}} - \overline{\delta\mathcal{E}}_{\text{odd}}|, \tag{4b}$$

where

$$\overline{\delta\mathcal{E}}_{\text{even}} = \left(\sum_{p=0}^{\text{NUPING}-1} \delta\mathcal{E}_{n-2p}\right)/\text{NUPING}, \tag{5a}$$

$$\overline{\delta\mathcal{E}}_{\text{odd}} = \left(\sum_{p=0}^{\text{NUPING}-1} \delta\mathcal{E}_{n-2p-1}\right)/\text{NUPING}, \tag{5b}$$

and

$$\Delta(\delta\mathcal{E})_{\text{even}} = \left(\sum_{p=0}^{\text{NUPING}-1} \left(\delta\mathcal{E}_{n-2p} - \overline{\delta\mathcal{E}}_{\text{even}}\right)^2\right)^{1/2}/\text{NUPING}, \tag{6a}$$

$$\Delta(\delta\mathcal{E})_{\text{odd}} = \left(\sum_{p=0}^{\text{NUPING}-1} \left(\delta\mathcal{E}_{n-2p-1} - \overline{\delta\mathcal{E}}_{\text{odd}}\right)^2\right)^{1/2}/\text{NUPING}. \tag{6b}$$

Here, $n$ denotes the number of the last accomplished HF iteration, and $\delta\mathcal{E}_{n'}$ denotes the stability energy (I-37) obtained in the $n'$-th iteration.

The default value of EPSPNG=0.0 ensures that whenever this new option is not used, the code HFODD (v1.75r) behaves as that in version (v1.60r). If a non-zero value of EPSPNG is used, a value of NUPING>1 has to be used too. In practice, values of EPSPNG=0.01 and NUPING=3 allow for an efficient detection of the "ping-pong" divergence condition.

Upon discovering the "ping-pong" divergence, the HF iterations are terminated and a table of absolute values of maximum differences of single-particle observables between the two sequences of iterations is printed, see Sec. 2.6 and Table 1. These maximum differences are determined for states in each of the charge–parity–signature (or charge–simplex) blocks, and separately for particle and hole states. Whenever such a maximum difference is found for a particle state and for a hole state with adjacent indices, such a pair is proposed as a candidate for the diabatic blocking calculation, see Secs. 2.6 and 3.2.

*Keyword:* CHAOTIC

$$0 = \text{NUCHAO}$$

The code is able to detect the "chaotic" divergence which occurs when the HF iterations give results which chaotically vary from one iteration to another one. The code recognizes such a divergence by finding the local maxima $M_k$, $k$=1,2,..., in the sequence of absolute values of the stability energies (I-37), obtained in the entire series of the HF iterations performed. The "chaotic" divergence condition occurs when the code finds NUCHAO positive differences $M_k - M_{k-1}$. When this condition occurs, iteration procedure terminates and the final results are printed.

For NUCHAO=0 (the default value) the code does not check whether the "chaotic" divergence occurs or not. In practice, a value of NUCHAO=5 allows for an efficient detection of the "chaotic" divergence condition. However, for a small value of NUCHAO and a small value of EPSPNG, the "ping-pong" divergence can sometimes be mistaken for the "chaotic" divergence. If one is interested in the diabatic-blocking data, printed after the "ping-pong" divergence, the recommended value of NUCHAO=5 should be increased to 10 or more.

*Keyword:* PHASESPACE

$$0, 0, 0, 0 = \text{NUMBSP(0,0), NUMBSP(1,0),}$$
$$\text{NUMBSP(0,1), NUMBSP(1,1)}$$

Numbers of the lowest mean-field eigenstates which are kept after the diagonalization of the mean-field Hamiltonians in the four simplex-charge blocks: $(s,q)=(+i,n)$, $(-i,n)$, $(+i,p)$, $(-i,p)$. All other eigenstates are discarded. If any of these numbers is equal to zero (the default value), the code sets it equal to the number of neutrons IN_FIX (for $q$=$n$) or protons IZ_FIX (for $q$=$p$), see Section 3.1 of II.

For calculations without pairing, the user is responsible for using values of NUMBSP large enough to accommodate all wave functions which might be useful for the required vacuum and particle-hole configurations, see Section 3.4 of II. In practice, the use of the default values described above is recommended as a safe option. The execution time is almost independent of NUMBSP. The size of the matrices defined by the NDSTAT parameter can be reduced by the user for smaller values of NUMBSP.

## 3.2 Configurations

*Keyword:* `DIASIM_NEU`

```
              2, 2,   1, 1,    0, 0 =   KPFLIM(0,0),KPFLIM(1,0),
                                        KHFLIM(0,0),KHFLIM(1,0),
                                        KOFLIM(0,0),KOFLIM(1,0)
```

Diabatic blocking of neutron single-particle simplex configurations. Matrices `KPFLIM` contain the indices of the particle states in the two simplex blocks denoted by (+) and (−), of given simplex values, i.e., $s=+i$ and $-i$, respectively. Matrices `KHFLIM` contain analogous indices of the hole states, and matrices `KOFLIM` define type of blocking according to the following table:

| | | |
|---|---|---|
| `KOFLIM=0`  | $\Longleftrightarrow$ | No diabatic blocking in the given parity/signature block. |
| `KOFLIM=+1` | $\Longleftrightarrow$ | The state which has the *larger* alignment is occupied. |
| `KOFLIM=−1` | $\Longleftrightarrow$ | The state which has the *smaller* alignment is occupied. |
| `KOFLIM=+2` | $\Longleftrightarrow$ | The state which has the *larger* intrinsic spin is occupied. |
| `KOFLIM=−2` | $\Longleftrightarrow$ | The state which has the *smaller* intrinsic spin is occupied. |

Within the diabatic blocking procedure one does not predefine whether the particle or the hole state is occupied (like is the case when the particle-hole excitations are defined, see Section 3.4 of II). In each iteration the code calculates the average alignments (or average intrinsic spins) of both states (those defined by `KPFLIM` and `KHFLIM`), and occupies that state for which a larger, or a smaller value is obtained. Therefore, the order of both states in the Routhian spectrum is irrelevant.

The user is responsible for choosing the particle-state indices (in `KPFLIM`) only among those corresponding to empty single-particle states, and the hole-state indices (in `KHFLIM`) only among those corresponding to occupied single-particle states, see Section 3.4 of II.

*Keyword:* `DIASIM_PRO`

```
              2, 2,   1, 1,    0, 0 =   KPFLIM(0,1),KPFLIM(1,1),
                                        KHFLIM(0,1),KHFLIM(1,1),
                                        KOFLIM(0,1),KOFLIM(1,1)
```

Same as above but for the diabatic blocking of proton single-particle simplex configurations.

*Keyword:* `DIASIG_NEU`

```
              2, 2, 2, 2,   1, 1, 1, 1,    0, 0, 0, 0 =
              KPFLIG(0,0,0),KPFLIG(0,1,0),KPFLIG(1,0,0),KPFLIG(1,1,0),
              KHFLIG(0,0,0),KHFLIG(0,1,0),KHFLIG(1,0,0),KHFLIG(1,1,0),
              KOFLIG(0,0,0),KOFLIG(0,1,0),KOFLIG(1,0,0),KOFLIG(1,1,0)
```

Diabatic blocking of neutron single-particle parity/signature configurations. Matrices `KPFLIG` contain the indices of particle states in the four parity/signature blocks denoted by (+,+), (+,−), (−,+), and (−,−), of given (parity,signature) combinations, i.e., $(\pi,r)=(+1,+i)$, $(+1,-i)$, $(-1,+i)$, and $(-1,-i)$, respectively. Matrices `KHFLIG` contain analogous indices of hole states, and matrices `KOFLIG` define the type of blocking according to the table of values identical to that defined for the simplex case above. Other rules described for the simplex case apply here analogously.

*Keyword:* `DIASIG_PRO`

                    2, 2, 2, 2,    1, 1, 1, 1,    0, 0, 0, 0 =
                    KPFLIG(0,0,1),KPFLIG(0,1,1),KPFLIG(1,0,1),KPFLIG(1,1,1),
                    KHFLIG(0,0,1),KHFLIG(0,1,1),KHFLIG(1,0,1),KHFLIG(1,1,1),
                    KOFLIG(0,0,1),KOFLIG(0,1,1),KOFLIG(1,0,1),KOFLIG(1,1,1)

Same as above but for the diabatic blocking of proton single-particle parity/signature configurations.

### 3.3 Files

*Keyword:* `COULOMFILE`

                    HFODD.COU = FILCOU

`CHARACTER*68` file name of the file containing auxiliary data for a faster calculation of the Coulomb potential. Must start at the 13-th column of the data line. Parameters `NUMCOU`, `NUMETA`, and `FURMAX` which define the calculations of the direct Coulomb potential (see Section 5 of I and Section 3.5 of II) are usually kept unchanged for the whole series of calculations in one region of nuclei. Therefore, many Coulomb auxiliary results can be calculated only once, and stored in the file `FILCOU`. Handling of this file is determined by the input parameters `ICOULI` and `ICOULO`.

*Keyword:* `COULOMSAVE`

                    0,0 = ICOULI, ICOULO

Input parameters `ICOULI` and `ICOULO` determine actions pertaining to reading and/or writing of the Coulomb file `FILCOU`, according to the following table:

| ICOULI | ICOULO | Action |
|:---:|:---:|:---|
| 0 | 0 | neither read nor write the Coulomb file |
| 1 | 0 | read, but do not write the Coulomb file |
| 0 | 1 | do not read, but write the Coulomb file |
| 1 | 1 | use automated handling of the Coulomb file |

The default values of `ICOULI`=`ICOULO`=0 ensure that whenever this new option is not used, the code HFODD (v1.75r) behaves as that in version (v1.60r). However, unless it is required by special circumstances, a use of the automated handling of the Coulomb file (`ICOULI`=`ICOULO`=1) is recommended. Within the automated mode, the code checks whether the Coulomb file exists, and whether it contains data which match the current values of the input parameters `NUMCOU`, `NUMETA`, and `FURMAX`. If this is the case, the code reads the data from the Coulomb file. If this is not the case, the code calculates the Coulomb auxiliary results and stores them in the Coulomb file `FILCOU`. In the automated mode, the user is informed by appropriate messages printed on the output file about what type of the action has been taken in the given run of the code.

## 4 Output file

Together with the FORTRAN source code in the file `hfodd.f`, an example of the output file is provided in `dy152-d.out`. Selected lines from this file are presented in the section

TEST RUN OUTPUT below. This output file corresponds to the input file `dy152-d.dat` reproduced in the section TEST RUN INPUT below.

## 5  FORTRAN source file

The FORTRAN source code in is provided in the file `hfodd.f` and can be modified in several places, as described in this section.

### 5.1  Library subroutines

The code HFODD requires an external subroutine which diagonalizes complex hermitian matrices. Version (v1.75r) contains an interface to the LAPACK subroutine ZHPEV that can be downloaded (with dependencies) from
http://netlib2.cs.utk.edu/cgi-bin/netlibfiles.pl?filename=/lapack/complex16/zhpev.f
This subroutine and its dependencies are in the `REAL*8` and `COMPLEX*16` versions, and should be compiled without promoting real numbers to the double precision. On the other hand, the code HFODD itself does require compilation with an option promoting to double precision. Therefore, the code and the ZHPEV package should be compiled separately, and then should be linked together.

In order to activate the interface to the LAPACK ZHPEV subroutine, the following modifications of the code HFODD (v1.75r) have to be made:

1. Change everywhere the value of parameter `I_CRAY=1` into `I_CRAY=0`.

2. Change everywhere the value of parameter `IZHPEV=0` into `IZHPEV=1`.

3. Change the name of the subroutine ZHPEV, provided with the code HFODD, to another name, or remove it from the file.

4. If your compiler does not support undefined externals, or subroutines called with different parameters, remove calls to subroutines CGEMM, F02AXE, *and the first call* to ZHPEV.

## 6  Acknowledgments

Useful comments and tests of the code performed by M. Bender, H. Molique, W. Satuła, and T.R. Werner are gratefully acknowledged. This research was supported in part by the Polish Committee for Scientific Research (KBN) under Contract No. 2 P03B 040 14, by the French-Polish integrated actions programme POLONIUM, and by the computational grant from the Interdisciplinary Centre for Mathematical and Computational Modeling (ICM) of the Warsaw University.

## References

[1] J. Dobaczewski and J. Dudek, Comput. Phys. Commun. **102**, 166 (1997).

[2] J. Dobaczewski and J. Dudek, Comput. Phys. Commun. **102**, 183 (1997).

[3] J. Dobaczewski and J. Dudek, Phys. Rev. **C52**, 1827 (1995).

[4] J.D. Jackson, *Classical Electrodynamics* (Wiley, New York, 1975).

[5] P. Ring and P. Schuck, *The Nuclear Many-Body Problem* (Springer-Verlag, Berlin, 1980).

[6] N. El Aouad, J.Dobaczewski, J. Dudek, X. Li, W.D. Luo, H. Molique, A. Bouguettoucha, Th. Byrski, F.A. Beck, D. Curien, G. Duchêne, Ch. Finck, and B. Kharraja, submitted to Nuclear Physics A.

# TEST RUN INPUT

```
==========================================================================
| This file (dy152-d.dat) contains the input data for the code HFODD.   |
|    Only keywords introduced after version (1.60r) are included here.   |
==========================================================================
                         ---------- General data  ----------
ITERAT_EPS
          0.0001
MAXANTIOSC
          5
PING-PONG
          0.01    3
CHAOTIC
          5
PHASESPACE
          0      0       0      0
                         ---------- Configurations --------
DIASIM_NEU              PARTICLES            HOLES              TYPE
                 00    00                00   00              0   0
DIASIM_PRO              PARTICLES            HOLES              TYPE
                 00    00                00   00              0   0
DIASIG_NEU              PARTICLES            HOLES              TYPE
           00    00    00    00        00    00   00   00    0  0  0  0
DIASIG_PRO              PARTICLES            HOLES              TYPE
           00    00    00    00        00    00   00   00    0  0  0  0
                         ------------- Files --------------
COULOMFILE
          HFODD.COU
COULOMSAVE
          1      1
                         ---------- Calculate -------------
EXECUTE
                         ---------- Terminate -------------
ALL_DONE
```

# TEST RUN OUTPUT

```
****************************************************************************
*                                                                          *
*   HFODD     HFODD     HFODD     HFODD     HFODD     HFODD     HFODD      *
*                                                                          *
****************************************************************************
*                                                                          *
*            SKYRME-HARTREE-FOCK CODE VERSION 1.75R                        *
*            ONE SYMMETRY-PLANE AND NO TIME-REVERSAL SYMMETRY              *
*            DEFORMED CARTESIAN HARMONIC-OSCILLATOR BASIS                  *
*                                                                          *
****************************************************************************
*                                                                          *
*                JACEK DOBACZEWSKI AND JERZY DUDEK                         *
*                INSTITUT DE RECHERCHES SUBATOMIQUES                       *
*              (FORMER CENTRE DE RECHERCHES NUCLEAIRES)                    *
*                     STRASBOURG, 1993-2000                                *
*                                                                          *
****************************************************************************

****************************************************************************
* PARAMETER SET SKM*:  T0= -2645.00  T1=   410.00  T2=  -135.00  T3= 15595.00 *
* POWER=0.1667  W=130  X0=  0.09000  X1=  0.00000  X2=  0.00000  X3=  0.00000 *
****************************************************************************

****************************************************************************
*                                                                          *
*   AUTOMATED HANDLING OF THE COULOMB FILE                                 *
*                                                                          *
****************************************************************************
*                                                                          *
*   THE COULOMB FILE WRITTEN ON THE DISC, FILE NAME (FILCOU):              *
*   HFODD.COU                                                              *
****************************************************************************
*                                                                          *
* NUCLIDE:  N = 86  Z = 66    NUMBER OF ITERATIONS =  50   CONTINUATION? = 0 *
*                                                                          *
****************************************************************************
*                                                                          *
* ITERATIONS WILL STOP WHEN THE STABILITY BECOMES LOWER THAN EPSITE=0.000100 *
*                                        FOR THE 5 CONSECUTIVE ITERATIONS  *
*                                                                          *
****************************************************************************
*                                                                          *
* ITERATIONS WILL STOP WHEN THE PING-PONG DIVERGENCE OCCURS FOR EPSPNG=0.010 *
*                                        DURING 3 PAIRS OF CONSECUTIVE ITERATIONS *
*                                                                          *
****************************************************************************
*                                                                          *
* ITERATIONS WILL STOP WHEN THE CHAOTIC-DIVERGENCE CONDITION OCCURS  5 TIMES *
*                                                                          *
****************************************************************************
*                                                                          *
* SLOW-DOWN PARAMETERS:      TIME-EVEN FIELDS = 0.50   TIME-ODD FIELDS = 0.50 *
*                                                                          *
****************************************************************************
*                                                                          *
* COULOMB PARAMETERS:   NUMCOU = 80  NUMETA = 79  BOUCOU = 20.0  FURMAX =0.25 *
*                                                                          *
****************************************************************************
*                                                                          *
* MAXIMUM NUMBER OF MULTIPOLES CONSIDERED: FOR THE CONSTRAINTS,   NMUCON = 2 *
*                                          FOR THE COULOMB FIELD, NMUCOU = 4 *
*                                          FOR THE OUTPUT INFO,   NMUPRI = 4 *
*                                                                          *
****************************************************************************
*                                                                          *
* MULTIPOLE CONSTRAINTS: LAMBDA= 2 MIU= 0  MOMENT= 42.000  STIFFNESS=0.010  *
*                                                                          *
****************************************************************************
*                                                                          *
* PRINTING THE RESULTS FOR THE FOLLOWING ITERATIONS:        THE FIRST  (1) *
*                                                          THE MIDDLE (0) *
*                                                    AND/OR THE LAST   (1) *
*                                                                          *
****************************************************************************
*                                                                          *
* NUMBERS OF LEVELS IN TIME-REVERSAL/CHARGE  BLOCKS:      TIME-UP  TIME-DOWN *
*                                           NEUTRONS:        86        86  *
*                                           PROTONS:         66        66  *
*                                                                          *
****************************************************************************
*                                                                          *
* CALCULATIONS WITHOUT PAIRING                                             *
*                                                                          *
****************************************************************************
*                                                                          *
* CALCULATIONS WITH PARITY/SIGNATURE SYMMETRY                              *
*                                                                          *
****************************************************************************
*                                                                          *
* CALCULATIONS WITH BROKEN TIME-REVERSAL SYMMETRY                          *
*                                                                          *
****************************************************************************
*                                                                          *
* INITIAL ROTATIONAL FREQUENCY OMEGAY = 0.500000                           *
*                                                                          *
****************************************************************************
*                                                                          *
* LINEAR CONSTRAINT ON SPIN                                                *
*                                                                          *
****************************************************************************
```

```
****************************************************************************
*                                                                          *
*   PARITY/SIGNATURE CONFIGURATIONS:                                       *
*                                                                          *
*                 V A C U U M        P A R T I C L E S         H O L E S  *
*              ===============      =================        =========    *
*              (++) (+-) (-+) (--)  (++) (+-) (-+) (--)  (++) (+-) (-+) (--) *
*                                                                          *
*   NEUTRONS:   22   22   21   21    0    0    0    0    0    0    0    0 *
*   PROTONS :   16   16   17   17    0    0    0    0    0    0    0    0 *
*                                                                          *
****************************************************************************
****************************************************************************
*                                                                          *
*   CONVERGENCE REPORT                                                     *
*                                                                          *
****************************************************************************
*                                                                          *
*    ITER     ENERGY      STABILITY      Q20    GAMMA    SPIN   OMEGA  HOW NICE *
*                                                                          *
*      0   -558.029407  -761.683371   54.178  -0.069   98.777  0.500  0.422842 *
*      1  -1163.372189    87.725963   48.776   0.047   52.033  0.500  1.081557 *
*      2  -1195.980005    63.414957   47.722  -0.025   49.155  0.500  1.055992 *
*      3  -1204.228467    38.471958   46.175   0.003   45.243  0.500  1.033002 *
*      4  -1207.038466    24.979693   45.544   0.020   45.509  0.500  1.021132 *
*      5  -1208.304258    15.158481   44.918   0.038   45.757  0.500  1.012705 *
*      6  -1208.924216     9.559571   44.422   0.055   46.049  0.500  1.007971 *
*      7  -1209.227387     6.030570   44.007   0.071   46.339  0.500  1.005012 *
*      8  -1209.364247     3.884991   43.664   0.086   46.612  0.500  1.003223 *
*      9  -1209.413569     2.543565   43.379   0.101   46.858  0.500  1.002108 *
*     10  -1209.417195     1.696859   43.142   0.114   47.074  0.500  1.001405 *
*     25  -1209.092516     0.016427   42.016   0.215   48.199  0.500  1.000014 *
*     26  -1209.085246     0.012900   42.002   0.217   48.214  0.500  1.000011 *
*     27  -1209.079025     0.010203   41.990   0.220   48.227  0.500  1.000008 *
*     28  -1209.073706     0.008125   41.980   0.222   48.238  0.500  1.000007 *
*     29  -1209.069161     0.006513   41.972   0.223   48.248  0.500  1.000005 *
*     30  -1209.065279     0.005253   41.965   0.225   48.256  0.500  1.000004 *
*     31  -1209.061965     0.004261   41.959   0.226   48.263  0.500  1.000004 *
*     32  -1209.059136     0.003474   41.954   0.228   48.268  0.500  1.000003 *
*     33  -1209.056722     0.002846   41.950   0.229   48.273  0.500  1.000002 *
*     34  -1209.054662     0.002342   41.946   0.230   48.278  0.500  1.000002 *
*     35  -1209.052904     0.001934   41.943   0.231   48.281  0.500  1.000002 *
*     36  -1209.051404     0.001603   41.941   0.231   48.284  0.500  1.000001 *
*     37  -1209.050124     0.001333   41.938   0.232   48.287  0.500  1.000001 *
*     38  -1209.049031     0.001112   41.937   0.233   48.289  0.500  1.000001 *
*     39  -1209.048098     0.000929   41.935   0.233   48.291  0.500  1.000001 *
*     40  -1209.047301     0.000778   41.934   0.234   48.292  0.500  1.000001 *
*     41  -1209.046621     0.000653   41.933   0.234   48.294  0.500  1.000001 *
*     42  -1209.046040     0.000549   41.932   0.235   48.295  0.500  1.000000 *
*     43  -1209.045544     0.000462   41.931   0.235   48.296  0.500  1.000000 *
*     44  -1209.045120     0.000389   41.931   0.235   48.297  0.500  1.000000 *
*     45  -1209.044757     0.000329   41.930   0.235   48.298  0.500  1.000000 *
*     46  -1209.044447     0.000278   41.930   0.236   48.298  0.500  1.000000 *
*     47  -1209.044182     0.000235   41.929   0.236   48.299  0.500  1.000000 *
*     48  -1209.043955     0.000199   41.929   0.236   48.299  0.500  1.000000 *
*     49  -1209.043761     0.000168   41.929   0.236   48.300  0.500  1.000000 *
*                                                                          *
****************************************************************************
****************************************************************************
*                                                                          *
*              DENSITY INTEGRALS IN THE SKYRME FUNCTIONAL                  *
*                                                                          *
****************************************************************************
*                                                                          *
*                TOTAL(T)       SUM(S)      ISOSCALAR(P)   ISOVECTOR(M)    *
*                --------       ------      ------------   ------------    *
*   DRHO_  =    17.558301      8.908547      17.558301       0.258793      *
*   DRHOD =    12.481232      6.331119      12.481232       0.181006      *
*   DLPR_ =    -3.718467     -1.885747      -3.718467      -0.053027      *
*   DTAU_ =    15.576998      7.972052      15.576998       0.367107      *
*   DSCU_ =     0.119021      0.063593       0.119021       0.008164      *
*   DDIV_ =     0.821038      0.417814       0.821038       0.014590      *
*                                                                          *
*   DSPI_ =     0.009448      0.005856       0.009448       0.002265      *
*   DSPID =     0.006720      0.004158       0.006720       0.001595      *
*   DLPS_ =    -0.004751     -0.006086      -0.004751      -0.007421      *
*   DCUR_ =     0.058252      0.030270       0.058252       0.002287      *
*   DKIS_ =     0.014893      0.007474       0.014893       0.000055      *
*   DROT_ =     0.004337      0.002462       0.004337       0.000587      *
*                                                                          *
****************************************************************************
****************************************************************************
*                                                                          *
*           CONTRIBUTIONS TO ENERGY IN THE SKYRME FUNCTIONAL               *
*                                                                          *
****************************************************************************
*                                                                          *
*                TOTAL(T)       SUM(S)      ISOSCALAR(P)   ISOVECTOR(M)    *
*                --------       ------      ------------   ------------    *
*   ERHO_  = -24265.791099   6951.116415  -17415.639545     100.964861     *
*   ERHOD =  16220.401153  -4113.908512   12165.300865     -58.808223     *
*   ELPR_ =    317.231713    -64.527908     253.611067      -0.907262     *
*   ETAU_ =   1070.918626   -543.096073     540.327125     -12.504572     *
*   ESCU_ =      0.000000      4.332261       4.054159       0.278102     *
*   EDIV_ =    -53.367500    -27.157931     -80.051250      -0.474181     *
*              ============  ============   ============   ============    *
*   SUM EVEN: -6710.607107   2206.758253   -4532.397579      28.548725    *
*                                                                          *
*   ESPI_ =     -0.562245      3.872319       2.561340       0.748734     *
*   ESPID =      0.000000     -2.701576      -2.183390      -0.518186     *
*   ELPS_ =      0.000000     -0.208253      -0.081291      -0.126962     *
*   ECUR_ =     -4.004811      2.062118      -2.020609       0.077916     *
*   EKIS_ =      0.000000     -0.509167      -0.507277      -0.001890     *
*   EROT_ =     -0.281919     -0.160046      -0.422878      -0.019087     *
*              ============  ============   ============   ============    *
*   SUM  ODD:   -4.848975      2.355395      -2.654105       0.160525     *
*                                                                          *
****************************************************************************
```

```
*******************************************************************************
*                                                                             *
*   SINGLE-PARTICLE PROPERTIES: HARTREE-FOCK                         NEUTRONS  *
*                                                                             *
*******************************************************************************
*                                                                             *
*  NO)    ENERGY  (++,+-,-+,--)  | N,nz,lz,OMEG>    <P>       JY      SY     GFACT  *
*                                                                             *
*  75)  -11.812 ( 0, 0,20, 0) | 7, 7, 0, 1/2>   -100    3.125   0.328   0.105 *
*  76)  -11.744 ( 0, 0, 0,20) | 5, 3, 2, 3/2>   -100    0.297  -0.063  -0.211 *
*  77)  -11.742 ( 0, 0,21, 0) | 5, 3, 2, 3/2>   -100    1.567   0.110   0.071 *
*  78)  -11.527 (19, 0, 0, 0) | 4, 1, 1, 1/2>    100    0.130  -0.165  -1.268 *
*  79)  -11.320 ( 0,19, 0, 0) | 4, 1, 3, 5/2>    100   -0.152  -0.004   0.026 *
*  80)  -11.284 (20, 0, 0, 0) | 4, 1, 3, 5/2>    100    0.169  -0.109  -0.648 *
*  81)  -11.251 (21, 0, 0, 0) | 6, 5, 1, 1/2>    100    0.771  -0.167  -0.217 *
*  82)  -11.034 ( 0,20, 0, 0) | 4, 1, 1, 1/2>    100    0.522  -0.004  -0.008 *
*  83)  -10.899 ( 0,21, 0, 0) | 4, 1, 1, 1/2>    100    0.345  -0.016  -0.046 *
*  84)  -10.416 (22, 0, 0, 0) | 6, 4, 2, 5/2>    100   -0.014  -0.031   2.177 *
*  85)  -10.405 ( 0,22, 0, 0) | 6, 4, 2, 5/2>    100    0.074  -0.019  -0.263 *
*  86)   -9.561 ( 0, 0, 0,21) | 7, 7, 0, 1/2>   -100    2.188  -0.036  -0.016 *
*  87)   -7.830 ( 0, 0,22, 0) | 5, 2, 1, 3/2>   -100    0.839   0.124   0.148 *
*  88)   -7.741 ( 0, 0, 0,22) | 5, 2, 1, 3/2>   -100    0.308   0.127   0.412 *
*  89)   -7.677 (23, 0, 0, 0) | 4, 0, 2, 5/2>    100   -0.282   0.087  -0.308 *
*  90)   -7.675 ( 0,23, 0, 0) | 4, 0, 2, 5/2>    100   -0.282   0.086  -0.304 *
*  91)   -7.521 ( 0, 0,23, 0) | 5, 2, 1, 3/2>   -100    0.975   0.087   0.089 *
*  92)   -7.215 ( 0, 0, 0,23) | 7, 7, 0, 1/2>   -100    0.026  -0.124  -4.839 *
*  93)   -7.208 ( 0, 0, 0,24) | 5, 1, 4, 9/2>   -100   -0.334   0.008  -0.024 *
*  94)   -7.208 ( 0, 0,24, 0) | 5, 1, 4, 9/2>   -100   -0.336   0.009  -0.026 *
*  95)   -6.936 ( 0,24, 0, 0) | 6, 4, 0, 1/2>    100    1.271   0.325   0.256 *
*  96)   -6.423 (24, 0, 0, 0) | 6, 4, 0, 1/2>    100    0.341  -0.415  -1.220 *
*  97)   -6.364 ( 0, 0, 0,25) | 5, 2, 1, 1/2>   -100    0.258  -0.215  -0.832 *
*  98)   -6.263 (25, 0, 0, 0) | 6, 3, 3, 7/2>    100   -0.015   0.009  -0.602 *
*  99)   -6.263 ( 0,25, 0, 0) | 6, 3, 3, 7/2>    100   -0.009   0.019  -2.112 *
* 100)   -6.164 ( 0,26, 0, 0) | 4, 0, 0, 1/2>    100    0.141   0.448   3.168 *
* 101)   -5.819 ( 0, 0,25, 0) | 5, 2, 1, 1/2>   -100   -0.045  -0.064   1.442 *
* 102)   -5.703 (26, 0, 0, 0) | 4, 0, 0, 1/2>    100   -0.575  -0.399   0.694 *
* 103)   -5.505 ( 0, 0,26, 0) | 5, 2, 3, 5/2>   -100    0.095  -0.021  -0.223 *
* 104)   -5.503 ( 0, 0, 0,26) | 5, 2, 3, 5/2>   -100    0.074  -0.012  -0.161 *
* 105)   -5.369 ( 0,27, 0, 0) | 6, 4, 2, 3/2>    100    0.350   0.004   0.011 *
* 106)   -5.366 (27, 0, 0, 0) | 6, 4, 2, 3/2>    100    0.608  -0.019  -0.032 *
* 107)   -5.172 ( 0,28, 0, 0) | 4, 0, 2, 3/2>    100   -0.519  -0.021   0.041 *
* 108)   -5.089 (28, 0, 0, 0) | 4, 0, 2, 3/2>    100   -0.445  -0.152   0.341 *
* 109)   -4.928 ( 0,29, 0, 0) | 4, 0, 4, 7/2>    100   -0.387  -0.031   0.080 *
* 110)   -4.926 (29, 0, 0, 0) | 4, 0, 4, 7/2>    100   -0.370  -0.040   0.109 *
* 111)   -4.793 ( 0,30, 0, 0) | 8, 8, 0, 1/2>    100    5.270   0.469   0.089 *
* 112)   -4.385 ( 0, 0, 0,27) | 7, 6, 1, 1/2>   -100    1.575  -0.105  -0.067 *
* 113)   -4.348 ( 0, 0,27, 0) | 7, 7, 0, 1/2>   -100    1.733   0.235   0.136 *
* 114)   -3.967 ( 0, 0,28, 0) | 7, 5, 2, 5/2>   -100    0.234  -0.038  -0.161 *
* 115)   -3.927 ( 0, 0, 0,28) | 7, 5, 2, 5/2>   -100   -0.157  -0.155   0.985 *
* 116)   -2.555 (30, 0, 0, 0) | 8, 7, 1, 3/2>    100    3.047   0.077   0.025 *
* 117)   -2.480 ( 0, 0, 0,29) | 5, 1, 2, 5/2>   -100   -0.074   0.067  -0.915 *
* 118)   -2.478 ( 0, 0,29, 0) | 5, 1, 2, 5/2>   -100   -0.067   0.069  -1.033 *
* 119)   -2.222 ( 0, 0, 0,30) | 5, 0, 5,11/2>   -100   -0.602  -0.002   0.003 *
*                                                                             *
*******************************************************************************
*                                                                             *
*   MULTIPOLE MOMENTS IN UNITS OF (10 FERMI)**LAMBDA                    TOTAL *
*                                                                             *
*******************************************************************************
*                                                                             *
*  Q00 =152.0000    .............  .............  .............  ............ *
*                                                                             *
*  Q10 =    ZERO  Q11 =    ZERO    .............  .............  ............ *
*                                                                             *
*  Q20 = 41.9283  Q21 =    ZERO  Q22 =  0.1730    .............  ............ *
*                                                                             *
*  Q30 =    ZERO  Q31 =    ZERO  Q32 =    ZERO  Q33 =    ZERO    ............ *
*                                                                             *
*  Q40 =  4.8357  Q41 =    ZERO  Q42 =  0.0074  Q43 =    ZERO  Q44 =-9.2E-04   *
*                                                                             *
*******************************************************************************
*                                                                             *
*   ROOT-MEAN-SQUARE AND GEOMETRIC SIZES IN FERMIS                      TOTAL *
*                                                                             *
*  R_RMS =  5.5449    X_RMS =  2.3841    Y_RMS =  2.3703    Z_RMS =  4.4095    *
*                                                                             *
*  R_GEO =  7.1584    X_GEO =  5.3310    Y_GEO =  5.3001    Z_GEO =  9.8599    *
*                                                                             *
*******************************************************************************
*                                                                             *
*   BOHR DEFORMATION PARAMETERS (FIRST-ORDER APPROXIMATION)             TOTAL *
*                                                                             *
*******************************************************************************
*                                                                             *
*  B10 =    ZERO  B11 =    ZERO    .............  .............  ............ *
*                                                                             *
*  B20 =  0.7112  B21 =    ZERO  B22 =  0.0029    .............  ............ *
*                                                                             *
*  B30 =    ZERO  B31 =    ZERO  B32 =    ZERO  B33 =    ZERO    ............ *
*                                                                             *
*  B40 =  0.5075  B41 =    ZERO  B42 =  0.0011  B43 =    ZERO  B44 =-1.4E-04   *
*                                                                             *
*******************************************************************************
```

```
****************************************************************************
*                                                                          *
*   ANGULAR MOMENTA AND THE FIRST MOMENTS OF INERTIA FOR OMEGA = 0.500000 MEV *
*                                                                          *
****************************************************************************
*                                                                          *
*                         SPINS                       J(1)                 *
*                ---------------------------  --------------------------   *
*                 ORBITAL   INTRINSIC   TOTAL    ORBITAL  INTRINSIC   TOTAL *
*                                                                          *
*   NEUTRONS     27.72295    0.90622  28.62917   55.44591   1.81243  57.25834 *
*   PROTONS      19.01301    0.65769  19.67069   38.02601   1.31537  39.34138 *
*   --------                                                                *
*   TOTAL        46.73596    1.56390  48.29986   93.47192   3.12780  96.59972 *
*                                                                          *
****************************************************************************
****************************************************************************
*                                                                          *
*                       NEUTRON   CONFIGURATIONS                           *
*                       ========================                           *
*        P S   12 13 14 15 16 17 18 19 20 21 22 23 24 25 26 27 28 29 30 31 32 *
*        ---   ---------------------------------------------------------    *
*                                                                          *
*  CONF: + +    1  1  1  1  1  1  1  1  1  1  0  0  0  0  0  0  0  0  0  0  0 *
*  VACC: + +    1  1  1  1  1  1  1  1  1  1  0  0  0  0  0  0  0  0  0  0  0 *
*                                                                          *
*  CONF: + -    1  1  1  1  1  1  1  1  1  1  0  0  0  0  0  0  0  0  0  0  0 *
*  VACC: + -    1  1  1  1  1  1  1  1  1  1  0  0  0  0  0  0  0  0  0  0  0 *
*                                                                          *
*  CONF: - +    1  1  1  1  1  1  1  1  1  1  0  0  0  0  0  0  0  0  0  0  0 *
*  VACC: - +    1  1  1  1  1  1  1  1  1  1  0  0  0  0  0  0  0  0  0  0  0 *
*                                                                          *
*  CONF: - -    1  1  1  1  1  1  1  1  1  0  0  0  0  0  0  0  0  0  0  0  0 *
*  VACC: - -    1  1  1  1  1  1  1  1  1  0  0  0  0  0  0  0  0  0  0  0  0 *
*                                                                          *
****************************************************************************
*                                                                          *
*                        PROTON   CONFIGURATIONS                           *
*                        ========================                          *
*        P S    7  8  9 10 11 12 13 14 15 16 17 18 19 20 21 22 23 24 25 26 27 *
*        ---   ---------------------------------------------------------    *
*                                                                          *
*  CONF: + +    1  1  1  1  1  1  1  1  1  0  0  0  0  0  0  0  0  0  0  0  0 *
*  VACC: + +    1  1  1  1  1  1  1  1  1  0  0  0  0  0  0  0  0  0  0  0  0 *
*                                                                          *
*  CONF: + -    1  1  1  1  1  1  1  1  1  0  0  0  0  0  0  0  0  0  0  0  0 *
*  VACC: + -    1  1  1  1  1  1  1  1  1  0  0  0  0  0  0  0  0  0  0  0  0 *
*                                                                          *
*  CONF: - +    1  1  1  1  1  1  1  1  1  1  0  0  0  0  0  0  0  0  0  0  0 *
*  VACC: - +    1  1  1  1  1  1  1  1  1  1  0  0  0  0  0  0  0  0  0  0  0 *
*                                                                          *
*  CONF: - -    1  1  1  1  1  1  1  1  1  0  0  0  0  0  0  0  0  0  0  0  0 *
*  VACC: - -    1  1  1  1  1  1  1  1  1  0  0  0  0  0  0  0  0  0  0  0  0 *
*                                                                          *
****************************************************************************
*                                                                          *
*                           ENERGIES (MEV)                                 *
****************************************************************************
*                                                                          *
*  KINETIC: (NEU)=  1651.194254   (PRO)=  1107.180948   (TOT)=  2758.375202 *
*  SUM EPS: (NEU)= -2033.727026   (PRO)= -1132.746046   (TOT)= -3166.473072 *
*  PAIRING: (NEU)=     0.000000   (PRO)=     0.000000   (TOT)=     0.000000 *
*                                                                          *
*  COULOMB: (DIR)=   564.274457   (EXC)=   -25.350821   (TOT)=   538.923637 *
*                                                                          *
*  CONSTR. (MULT)=     0.000051   SLOPE=    0.001433   CORR.=    -0.030046 *
*  CONSTR. (SPIN)=   -24.149930   SLOPE=    0.500000   CORR.=   -12.074965 *
*  REARRANGEMENT ENERGY FROM THE SKYRME DENSITY-DEPENDENT TERMS=  1008.649255 *
*  ROUTHIAN  (TOTAL ENERGY PLUS MULTIPOLE AND SPIN CONSTRAINTS)= -1233.193474 *
*                                                                          *
*  SPIN-ORB (EVE)=   -80.525431   (ODD)=    -0.441965   (TOT)=   -80.967396 *
*  SKYRME:  (EVE)= -4503.848854   (ODD)=    -2.493580   (TOT)= -4506.342434 *
*                                                                          *
*  TOTAL:   (STAB)=    0.000143   (SP)= -1209.043452   (FUN)= -1209.043595 *
*                                                                          *
****************************************************************************
****************************************************************************
*                                                                          *
*  REAL-CLOCK  EXECUTION  TIMES IN SUBROUTINES                             *
*  INCLUDING THEIR COMPLETE DOWN-CALLING TREES                             *
*                                                                          *
****************************************************************************
*                                                                          *
*      1023 => HFODD        0 => GEOMFC       21 => POWALL       0 => RECOUL *
*                                                                          *
*         1 => NILSON       3 => INTKIN      180 => DIASIG     174 => DIAMAT *
*                                                                          *
*         3 => AVPARI      96 => NILABS       64 => AVIMRE      37 => INTMUL *
*                                                                          *
*       261 => DENSHF      64 => DENMAT       37 => MOMETS      50 => BEGINT *
*                                                                          *
*        94 => INTCOU      16 => SKFILD        7 => RECORD     153 => INTEGH *
*                                                                          *
*        24 => INTMAS      12 => INTCEN       45 => INTSOR       8 => ANGYSP *
*                                                                          *
****************************************************************************
```